\documentclass[12pt]{iopart}
\usepackage[dvips]{graphicx}
\usepackage{amssymb}
\usepackage{amsfonts}
\usepackage{multirow}
\usepackage{color}
\usepackage{colortbl}
\usepackage[utf8]{inputenc}
\usepackage{arydshln}
\usepackage{cite}

\newcommand{\gp}{\ensuremath{g'}} 
\newcommand{\lp}{\ensuremath{l'}}

\newcommand{\psic}{\ensuremath{\psi^\mathrm{c}}}
\newcommand{\eps}{\ensuremath{k}}
\newcommand{\abs}[1]{\ensuremath{\left| #1 \right|}}
\renewcommand{\vec}[1]{\ensuremath{{\textbf{\em #1}}}}

\newcommand{\pardiff}[2]{\ensuremath{\frac{{\rm \partial} #1}{{\rm \partial #2}}}}

\renewcommand{\Im}[1]{\ensuremath{{\rm Im}\left(#1\right)}}
\newcommand{\W}{\phantom{+}}

\begin{document}
\title[Mesoscale induced dynamical instabilities]{Engineering mesoscale structures with distinct dynamical implications in networks of delay-coupled delay oscillators}

\author{Anne-Ly Do$^1$, Johannes H\"ofener$^1$, and Thilo Gross$^2$}

\address{$^1$Max-Planck-Institute for the Physics of Complex Systems, Dresden, Germany\\
$^2$University of Bristol, Department of Engineering Mathematics, Bristol, UK}
\ead{ly@pks.mpg.de}

\begin{abstract} 
The dynamics of networks of interacting systems depends intricately on the interaction topology.  
When the dynamics is explored, generally the whole topology has to be considered. 
However, we show that there are certain mesoscale subgraphs that have precise and distinct consequences for the system-level dynamics. 
In particular, if mesoscale symmetries are present then eigenvectors of the Jacobian localise on the symmetric subgraph and the corresponding eigenvalues become insensitive to the topology outside the subgraph. 
Hence, dynamical instabilities associated with these eigenvalues can be analysed without considering the topology outside the subgraph. 
While such instabilities are thus generated entirely in small network subgraphs, they generally do not remain confined to the subgraph once the instability sets in and thus have system-level consequences. 
Here we illustrate the analytical investigation of such instabilities in an ecological meta-population model consisting of a network of delay-coupled delay oscillators.  
\end{abstract}
\submitto{\NJP}
\maketitle

\section{Introduction}
Over the past decade networks have become the principal tool for the analysis of complex systems \cite{Science}. 
By describing a given complex system as a network of discrete \emph{nodes} that interact via discrete \emph{links}, 
a considerable simplification is achieved, but the complexity of the topology of interactions (and thus much of the emergent properties of the real world system) is retained. 

In some of the biggest success stories of graph and network theory the simplification that is achieved by treating a system as a network, enables a subsequent analytical investigation that links an emergent phenomenon to local properties of the nodes. Examples include for instance Euler's solution of the K\"onigsberg bridge problem and the giant-component transition in random graphs \cite{Erdoes-Renyi}. 

It is apparent that not all emergent phenomena can be traced back to node properties alone. 
Instead, progress has been made by linking phenomena to well-studied global properties such as the spectra of matrices.
In particular, the spectra of the adjacency matrix and the different graph Laplacians contain information about the net's diameter \cite{Chung}, the degree of modularity \cite{McSherry} and the isoperimetry (i.e., how many edges have to be removed to cut it in pieces of a certain size) \cite{Chung}. Also in studies of the dynamics on networks clever mapping to a spectral problem has often proved instrumental. Prominent examples include for instance the computation of percolation thresholds of networks \cite{Boguna} and the master stability function approach to synchronization\cite{Pecora}. 

A direct connection between dynamics and spectral properties can be made for dynamical systems close to steady states \cite{Kuznetsov}. 
Here the relevant matrix is the Jacobian, which consists of the coefficients of a linearised system at the steady state.
In a given dynamical system steady state is stable if all eigenvalues of the Jacobian matrix have negative real parts.
Notably, this connection between the stability of steady states and the spectrum of the Jacobian holds in all dynamical systems including strongly heterogeneous ones. 

While linking emergent phenomena to spectral properties has led to considerable progress, it does not provide a satisfactory solution for the investigation of many real world networks, because the size of the network can be such that computation of the spectrum is infeasible and, more importantly, information on the network structure is often incomplete. 
Perhaps inspired by \cite{Alon}, this has triggered a recent search for mesoscale subgraphs that permit conclusions on the spectra of important matrices and by extension emergent phenomena. 

Compared to statistical null-models, the spectra of real world networks \cite{Goh,Kamp,Vukadinovic,Banerjee} display characteristic peaks, which can in part be linked to the statistical over-representation of certain subgraphs \cite{Golinelli,Vukadinovic,Banerjee,Kamp}. Thus, for instance, the multiplicity of the eigenvalue $0$ of a graph's Laplacian matrix has been found to increase with the number of star-like subgraphs with two or more leaf nodes \cite{Dorogovtsev,He,Goh}. 
A more applied approach was taken for instance in \cite{Bascompte} which studied the impact of three-node subgraphs on the stability of ecological food webs. 
 
Mathematically speaking, an eigenvalue of a network can be linked to a specific subgraph if the corresponding eigenvector $\vec{v}$ is \emph{localised} on the subgraph. 
In this case the eigenvector $\vec{v}$ and the corresponding localised eigenvalue ${\rm Ev}(\vec{v})$ are independent of the topology outside the subgraph \cite{Banerjee}. 
Thus, the localised eigenvector and the corresponding eigenvalue are preserved in all networks, in which the subgraph occurs.  

For a long time, the only well-studied examples for subgraphs with localised eigenvalues were star-like subgraphs, and complete subgraphs (cliques) \cite{Kamp}. Starting from the observation that spectral peaks are often associated with network symmetries \cite{Lauri}, 
MacArthur and co-workers determined generic symmetry properties that lead to the localization of eigenvectors \cite{McArthur, McArthur2008}. 
By describing these properties in terms of a net's automorphism group, they provided means to construct and classify arbitrarily large subgraphs with localised eigenvalues. 

The aim of the present paper and a companion paper \cite{Helge} is to apply the results of \cite{McArthur} to the study of dynamics and hence discuss the effect of subgraphs with localised eigenvectors for network dynamics. 
In the present paper we demonstrate these consequences by investigating a system of delay-coupled delay oscillators, which has been previously proposed as a model for the ecological metapopulations and gene-regulatory nets \cite{Johannes}. 
In contrast to the food web model studied in \cite{Helge}, the metapopulation model lends itself to a deeper analysis as it is a so-called \emph{factorising system} \cite{Johannes}.
In these systems the effect of topology can be `factored-out' such that eigenvalues of the Jacobian, which characterise the dynamics, can be written as eigenvalues of the adjacency matrix, which characterize the structure. 
While we exploit this property extensively to provide analytical results that illustrate our findings, we note that these findings are not limited to factorising systems. 
As the main result of the present paper, we show that it is possible to engineer subgraphs that have distinct dynamical consequences, irrespective of the embedding network.

This paper is organized as follows: 
We start in Sec.~\ref{BSS} by recalling the conditions under which a mesoscale structure has localised eigenvalues.
In Sec.~\ref{DelayOscillators}, we introduce a class of models of delay-coupled delay oscillators. 
Using the generalized modeling approach, we derive bifurcation conditions that explicitly depend on the topological eigenvalues.  
In Sec.~\ref{Numerics}, we study model realizations that include BSS and show that the localised eigenvalues give rise to distinct instabilities of the system. 
Finally, in Sec.~\ref{Conclusions}, we summarize and discuss our results.


\section{localised topological eigenvalues}\label{BSS}
The topology of a network of $N$ nodes can be described by an $N\times N$ matrix, the so-called \emph{adjacency matrix}. In case of a directed network, this matrix is defined as
\begin{equation}
	{\rm \bf A}: A_{ij}=\Biggl\{\begin{minipage}{7cm} 1 \textnormal{if there is a link pointing from $j$ to $i$} \\
	0  \textnormal{otherwise} 
	\end{minipage} \quad .
\end{equation}
In case of an undirected network, $A_{ij}=A_{ji}$.   

We say that a vector $\vec{v}\in \mathbb{C}^N$ is a topological eigenvector of a network if it is an eigenvector of the network's adjacency matrix ${\rm \bf A}$. 
Below, we discuss under which conditions topological eigenvectors can be attributed to a specific subgraph. 
The discussion presented here aims to provide an intuitive understanding. For a rigorous analysis we refer the reader to \cite{McArthur, McArthur2008}. 

\begin{figure}
\begin{minipage}{0.45\textwidth}
\begin{center}
\includegraphics{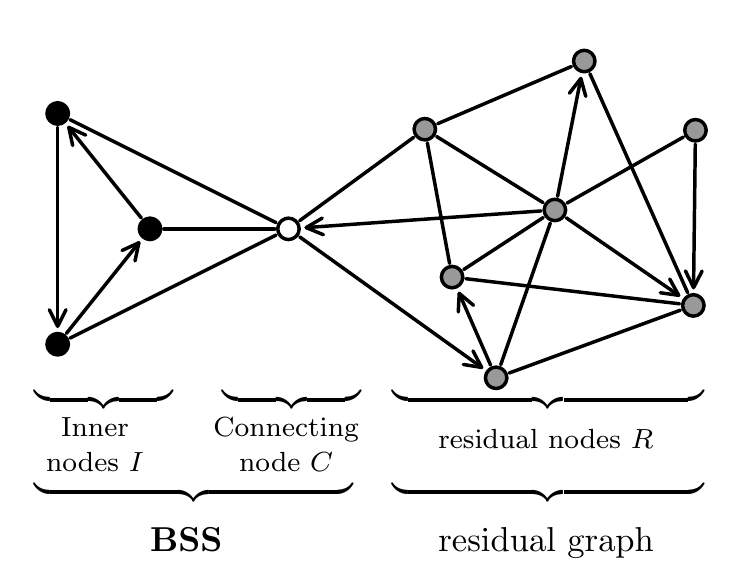}
\end{center}
\end{minipage}
\begin{minipage}{0.5\textwidth}
\caption{Decomposition of a network in a basic symmetric subgraph (BSS) and a residual graph. The BSS constitutes of inner nodes $I$ (black circles) and a connecting node (open circle). The residual graph constitutes of nodes $R$ (grey circles). Lines with arrows denote directed links, lines without arrows undirected links.} \label{fig:Decomp}
\end{minipage}
\end{figure}

Let us consider a small subgraph that is part of a bigger network. 
We group the nodes of the network into three sets: the set $I$ of inner nodes, which are in the subgraph and adjacent only to nodes of the subgraph; the set $C$ of connecting nodes, which are in the subgraph but not in $I$ which constitute the interface between the subgraph and the residual graph; and the set $R$ of residual nodes, which are not in the subgraph (cf.~Fig~\ref{fig:Decomp}).

Below we say that a topological eigenvector is localised on a given subgraph, if the components of the eigenvector vanish in all nodes except those belonging to the set $I$, i.e. $v_{j}=0\ \forall j\in R,C$. \footnote{We note that the notion of localization used here differs from the one in \cite{Helge}, where it was only required that the $\vec{v}$ vanishes on R. This difference is due to the present focus on engineering different functions in subgraphs, which requires an extra layer of `insulation' from the residual net.} The defining condition for a localised eigenvector is thus
\begin{equation}\label{R_cond}
\sum_{k}A_{jk}v_k=c\cdot v_j \Biggl\{\begin{minipage}{4cm}
 $= 0$ for all $j\in R,C$\\
 $\in \mathbb{C}$ for all $j\in I$ 
\end{minipage} \ .
\end{equation}

Let us discuss the implications of Eq.~\eref{R_cond}.
First, consider nodes $j \in R$. By definition, these nodes can only have neighbours $k\in R,C$. 
As $v_{k}=0\ \forall k\in R,C$, we can conclude that (i) Eq.~\eref{R_cond} is trivially fulfilled for $j \in R$, and that (ii) the values of all coefficients $A_{jk}$, $j\in R$ and $k\in R\cup C$ can be changed without changing the right hand side of any of the Eqs.~\eref{R_cond}. 
In other words: changes of the residual graph's topology, its size and/or its coupling to connecting nodes do not affect the eigenvector property of $\vec{v}$. 

Next, consider nodes $j \in C$, i.e., the connecting nodes.
These nodes have neighbours $\in R,C$ and $\in I$, which contribute to the sum in Eq.~\eref{R_cond}.
As $v_{k}=0$ for $k\in R,C$, contributions from all R-neighbours and C-neighbours of a C-node vanish. 
Thus, $\sum_{k}A_{jk}v_k=0$ requires that contributions from all I-neighbours of a C-node cancel each other. \footnote{This corresponds to the orbitwise cancelation criterion in \cite{McArthur}}.   

Finally, consider nodes $j\in I$. The respective entries $v_{j}$ can be non-zero but are subject to two conditions imposed by Eq.~\eref{R_cond}:
Firstly, as shown above, the first line of Eq.~\eref{R_cond} requires that there are entries $v_j$ $j\in I$ that sum up to zero.
Secondly, for the general case of complex $v_j=: p_j+ \imath q_j$, the second line of Eq.~\eref{R_cond} requires that $\sum_{k}A_{jk}(v_k-v_j)=\Im{c}$ for all $j\in I$, i.e., that the accumulated phase difference of node to its neighbours is identical for all nodes $j\in I$. 
Both conditions can be fulfilled if the $q_j$ are equally spaced on the unit circle, which is the case if 
the nodes $j\in I$ are subject to symmetry relations \cite{Spencer}. 

\begin{table}
\begin{center}
\begin{tabular}{ccl}
\hline
\hline\\
Subgraph & Eigenvalue & Eigenvector \\
\hline
\multirow{2}{*}{
\begin{minipage}{20mm}
\includegraphics[scale=0.55]{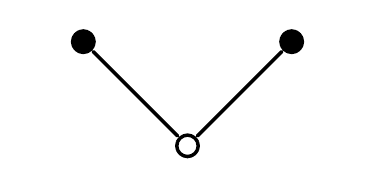}
\end{minipage}}
& $\pm\sqrt{2}$ & $(+,+,|\pm\sqrt{2})$\\
& $0$ & $(+,-,|0)^*$\\
\hline
\multirow{3}{*}{
\begin{minipage}{20mm}
\includegraphics[scale=0.55]{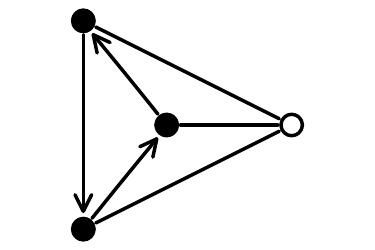}
\end{minipage}}
& $\xi_\pm$ & $(+,+,+,|3/\xi_\pm)$\\
& $\phi_1$ & $(1,\phi_2,\phi_1,|0)^*$\\
& $\phi_2$ & $(1,\phi_1,\phi_2,|0)^*$\\
\hline

\multirow{5}{*}{
\begin{minipage}{20mm}
\includegraphics[scale=0.55]{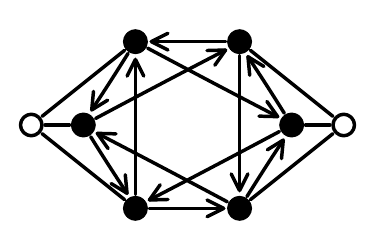}
\end{minipage}}
& $3$ & $(+,+,+,+,+,+,|1,1)$\\
& $\pm\sqrt{3}$ & $(+,+,+,-,-,-,|\pm\sqrt{3},\mp\sqrt{3})$\\
& $-1$ & $(-,-,-,-,-,-,|3,3)$\\
& $2\phi_1$ & $(1,\phi_2,\phi_1,1,\phi_2,\phi_1,|0,0)^*$\\
& $2\phi_2$ & $(1,\phi_1,\phi_2,1,\phi_1,\phi_2,|0,0)^*$\\
& $0$ & $(+,-,\W,-,+,\W,|0,0)^*$\\
& $0$ & $(+,\W,-,-,\W,+,|0,0)^*$\\
\hline
\hline
\end{tabular}
\end{center}\caption{Topological eigenvalues and eigenvectors of three BSS. Shown are a star-like subgraph, a directed triangle and a directed hexagon with one or two connecting nodes respectively. Inner nodes are denoted by closed circles and enumerated counter-clockwise beginning from the the upper left corner. Connecting nodes are denoted by open circles and enumerated last. Lines with arrows denote directed links, lines without arrows undirected links. We use the following abbreviations: $\phi_n:=\exp(2\pi n i/3)$, $\xi_\pm:=(1\pm\sqrt{13})/2$, and $\zeta_\pm:=(1\pm\sqrt{7})$; $+,-$ and the white-space denote $1,-1$ and $0$. localised eigenvectors, are distinguished by vanishing values of the eigenvector on the connecting nodes (after the `|'). Additionally we have marked the localised eigenvectors by a star ($^*$).
\label{Examples}}
\end{table}

Let us shortly summarize. The above reasoning showed that eigenvectors localised on a subgraph are retained in any network in which an arbitrary residual graph is attached in an arbitrary manner to the connecting node(s). Further, we motivated that localised eigenvectors occur on all subgraphs whose inner nodes are subject to symmetry relations. In the following we call those subgraphs basic symmetric subgraphs (BSS). Each BSS thus has one or more localised eigenvectors, whose corresponding eigenvalues appear in the spectrum of every network that contains the BSS. 

A classification of BSS with only undirected links and a recipe for their construction can be found in \cite{McArthur, McArthur2008}. 
Because a generalization of the scheme to directed BSS is straightforward, the present paper focusses on dynamical implications, rather than providing a wealth of examples for topological localization.
For illustration, we thus restrict ourselves to three generic BSS with both, directed and undirected links, which are shown in Table~\ref{Examples}. 


\section{Delay-coupled delay oscillators}~\label{DelayOscillators}
In the following, we present a generalized model of delay-coupled delay oscillators, which we use to study the dynamical implications of localised topological eigenvalues. The class of models has previously proposed in \cite{Johannes}, but is here extended to the case of directed links which has not previously been considered.

Delay networks constitute a powerful, and intensively studied framework for modelling biological systems \cite{Mackey1977,Ikeda,Glass,Losson1993,Giacomelli1996,Schanz2003,Wolfrum2006}. 
Modelling aspects of biology often requires to reduce complex processes to simple steps within a mechanism. 
Delay terms in models can account for time requirements of complex processes, such as transcription, gestation, etc., without resolving the process in detail \cite{Hirata2004}.
In the context of the present paper, delay networks offer also another advantage: 
Even small delay systems can exhibit complex dynamics, which allows us to study nontrivial dynamics in relatively small example systems \cite{Schoell}. 
In systems without delay, very similar results could be produced, but would require larger examples and a considerably more cumbersome notation. 

The analytical treatment of delay networks is generally demanding as even simple delay differential equations constitute infinite dimensional dynamical system and lead to transcendental equations \cite{Farmer1982}. We therefore focus on a class of delay-coupled delay networks on degree homogeneous networks, which are known to fall into a class of factorizing systems\cite{Johannes} and thus can be treated analytically. 

\subsection{The meta-population model}
We consider a directed network of $N$ nodes, corresponding to different patches that can be inhabited by a given species.
Each node $i$ has an internal dynamical variable $X_i$ representing the abundance of the population within the patch. In time the variable $X_i$ changes in time due to internal dynamics and due to coupling to other variables $X_j$ according to
\begin{equation}\label{eq:DGL}
	\dot{X}_i=G(X_i^{\tau})-L(X_i)+\sum_{j} \left[A_{ij}F(X_j^{\delta})-A_{ji}F(X_i)\right] 
\end{equation}
where $\tau$ and $\delta$ denote growth and travel-time delays, $A_{ij}$ and $A_{ji}$ denote entries of the adjacency matrix, and $G$, $L$ and $F$ are positive functions describing growth, loss, and migration, respectively. In the following we do not restrict the functions $G$, $L$, and $F$ to specific functional forms, but consider formally the whole class of models, using the approach of generalized modelling \cite{Gross,Gross2009,Gehrmann11}. The class of models under consideration thus includes several well-studied examples such as the Mackey-Glass \cite{Mackey1977} and the Ikeda model \cite{Ikeda}.

\subsection{Generalized stability analysis}
For simplicity we study Eq.~\eref{eq:DGL} on a network, in which every node has as many incoming links $d_i^\mathrm{in}$ as outgoing links $d_i^\mathrm{out}$, such that $d_i^\mathrm{in}=d_i^\mathrm{out}=d_i$. 
In this class of networks, which includes all undirected networks, flows tend to balance unequal node loads. Thus, if the functions $G$, $L$ and $F$ are identical for all nodes, there is homogeneous steady state, in which all variables $X_i^*=X^*$ are identical.

We analyse the stability of $X^*$ by introducing normalized variables $x_i = X_i/X^{*}$ and normalized functions $f(x_i) = F(x_iX^*)/F(X^*)$. Using $X^{*}= X^{\tau *}= X^{\delta *}$, we can rewrite Eq.~\eref{eq:DGL} as
\begin{equation}\label{eq:DDE_norm}
	x_{i}=\alpha(g(x_{i}^{\tau}))-l(x_i)+\beta\sum_{j}\left[A_{ij}f(x_j^{\delta})-A_{ji}f(x_i)\right]
\end{equation}
where $\alpha = G(X^*)/X^* = L(X^*)/X^*$ and $\beta= F(X^{*})/X^*$ are parameters that can be interpreted as characteristic turnover and transport rates. We set $\alpha = 1$ by timescale normalization.

The Jacobian matrix that governs the stability of steady states in the system can be expressed as a function of the quantities $g'$, $l'$ and $f'$ that are defined as derivatives of the normalized functions with respect to the normalized variable $s_i$, or equivalently the logarithmic derivatives of the functions before normalization
\begin{equation} \label{eqGenPar}
\gp=\left.\pardiff{g}{x}\right|_{x=1}=\left.\frac{X^*}{G^*}\pardiff{G}{X}\right|_{X=X^*}=\left.\pardiff{\log{G}}{\log{X}}\right|_{X=X^*}.
\end{equation}
Because we did not restrict the functional forms in the model the three quantities $g'$, $l'$ and $f'$ are unknown constants that can be treated as parameters of the system. 
Defining the parameters thus, after normalization, has the advantage of yielding parameters that are generally easily interpretable in the context of the application \cite{Gross}.

Following \cite{Johannes} we write the Jacobian as
\begin{equation}\label{eq:Jacobian}
{\rm \bf J}: J_{ij}(\lambda)=J_{i}^\mathrm{d}\delta_{ij}+J^\mathrm{o}A_{ij}\ ,
\end{equation}
where $\delta_{ij}$ is the Kronecker delta and
\begin{eqnarray}
J_i^\mathrm{d}&=(g'\exp(-\lambda\tau)-l')-d_i\beta f'\ ,\\
J^\mathrm{o}&=\beta f' \exp(-\lambda \delta)\ 
\end{eqnarray}
and the degree $d_i$, is the number of incoming and outgoing links connecting to node $i$ and $\lambda$ is an eigenvalue of the ${\rm  \bf J}$. 

\subsection{Bifurcation conditions}
In a continuous-time dynamical system, a steady state is locally asymptotically stable if all eigenvalues, ${\rm Ev}({\rm \bf J})$, have negative real-parts and it is unstable if at least one eigenvalue has a positive real part \cite{Kuznetsov,Atay}.
Because the Jacobian of a delay system contains exponentials of its own eigenvalue $\lambda$ the characteristic polynomial of the Jacobian is typically a transcendental equation that can admit an infinite number of eigenvalues. 

To provide some analytical intuition we now focus on the case of degree homogeneous networks, where $d_i=d^*$ for all $i$, which enables an analytical computation of the eigenvalues. For simplicity of notation we will for the moment pretend that degree homogeneity is satisfied for the entire network. We emphasize however that for the computation of localised eigenvalues and eigenvectors, we only need to assume degree homogeneity within the I-nodes, which is often true because of symmetry. Because the localised eigenvectors and eigenvalues are insensitive to the network structure outside the BSS, they are also insensitive to a violation of the degree homogeneity assumption if the violation occurs outside the BSS. It can be shown that essentially the same is true for the C-nodes.       

For degree-homogeneous networks all diagonal elements of the Jacobian are identical such that $J_i^\mathrm{d} =: J^\mathrm{d}$. We can then write 
\begin{equation}
{\rm \bf J} = J^\mathrm{d}{\rm \bf I} + J^\mathrm{o}{\rm \bf A}
\end{equation}
where ${\rm \bf I}$ is the identity matrix.
The eigenvalue computation then becomes 
\begin{equation}
\label{eqEigenvalueRel}
{\rm Ev}({\rm \bf J}) = {\rm Ev}(J^\mathrm{d}{\rm \bf I} + J^\mathrm{o}{\rm \bf A}) = J^\mathrm{d} + J^\mathrm{o} {\rm Ev}({rm \bf A})
\end{equation}
This equation establishes a relationship between the topological and dynamical eigenvalues of the system. However, we emphasize that both $J^\mathrm{d}$ and $J^\mathrm{o}$ depend on $\lambda$. The relationship (\ref{eqEigenvalueRel}) thus constitutes a self-consistency condition that typically admits an infinite set of dynamical eigenvalues for each topological eigenvalue. For a given topological eigenvalue $c_i$ this condition reads
\begin{equation}\label{eq:EV_DHON}
\lambda=J^\mathrm{d}(\lambda)+c_i J^\mathrm{o}(\lambda).
\end{equation} 

Instead of trying to compute the dynamical eigenvalues, we focus on the computation of bifurcation points where the stability changes and eigenvalues cross the imaginary axis in the complex plane, so that $\lambda=i \omega$. Separating the real and imaginary part of Eq.~\eref{eq:EV_DHON} yields
\begin{eqnarray}
\label{eq:Re}0&=\gp\cos(\phi)-\lp-d\eps+|c|\eps\cos(\psi),\\
\label{eq:Im}\omega&=-\gp\sin(\phi)-|c|\eps\sin(\psi),
\end{eqnarray}
with $\phi=\omega\tau$ and $\psi:=\omega\delta-\psic$, where $\psic$ is the complex phase of $c_i$.

The bifurcation condition is thus given by an equation system with three unknown variables, $\phi$, $\psi$ and $\omega$. For every solution triplet $(\phi,\psi,\omega)$, we find another solution $(-\phi,-\psi,-\omega)$, providing the same bifurcation lines. We therefore only consider solutions with $\omega>0$. Further, we are only interested in solutions for positive delays, $\tau>0$, and thus only consider solutions with $\phi>0$. We now choose $\phi$ as a free parameter to find parametric representations of the bifurcation lines. This is done separately for three cases $|c|=0$, $|c|=d$, and $0<|c|<d$. 

For $c=0$, Eqs.~(\ref{eq:Re},\ref{eq:Im}) are independent of $\psi$, $\delta$ respectively. Thus, the bifurcations are vertical lines in the $(\eps,\delta)$-plane, which are solely characterized by 
\begin{equation}\label{eq:beta_c0}
\eps=\frac{1}{\tau}\frac{h}{d},
\end{equation}
with $h:=(\gp\cos(\phi)-\lp)\tau$, where $\phi$ has to satisfy $f(\phi)=0$, with $f:=\phi+\gp\tau\sin(\phi)$. For the parameter values used throughout this paper, the condition $f(\phi)=0$ has a unique solution.

For $|c|=d$, we find
\begin{equation}\label{eq:beta_cd}
\eps=\frac{1}{\tau}\frac{f^2+h^2}{2dh}.
\end{equation}
In order to obtain positive solutions, $h(\phi)$ needs to be positive. Therefore, we have to restrict $\phi$ to the intervals $I^\eps_r=[2\pi r + \phi^\eps,2\pi(r+1)-\phi^\eps]$, where $r$ is a non-negative integer and $\phi^\eps=\cos^{-1}(\lp/\gp)$. 
The corresponding $\delta$ values can be calculated using the Eqs.~(\ref{eq:Re},\ref{eq:Im}), which yield
\begin{eqnarray}\label{eq:delta}
\delta^\mathrm{L,R}&=\frac{\psi^\mathrm{L,R}+\psi_{\rm c}+2\pi s}{\phi}\tau,\\
\psi^\mathrm{L,R}&=\pm\cos^{-1}\left(\frac{d}{|c|}+\frac{\lp-\gp \cos(\phi)}{|c|\eps}\right).
\end{eqnarray}
for non-negative integers $s$. Thus, after calculating $\eps$ with Eq.~\eref{eq:beta_cd}, we are able to calculate $\delta$. In order to obtain valid solutions, we need to apply the L-branch for all $f(\phi)<0$ and the R-branch otherwise.

Finally, for $0<|c|<d$, we find
\begin{equation}\label{eq:AD_beta}
\eps_\mathrm{1,2}=\frac{1}{\tau}\frac{dh\pm\sqrt{a}}{d^2-\abs{c}^2},
\end{equation}
with
\begin{equation}
\label{eq:a}a=d^2h^2-(d^2-|c|^2)(f^2+h^2).
\end{equation}
In order to obtain real and positive solutions, not only $h(\phi)$ needs to be larger than 0 but also $a(\phi)$. This sets additional restrictions to $\phi$, which can be calculated numerically. In contrast to the case $|c|=d$, valid solutions can only be found inside a finite number of intervals $I^\mathrm{\eps}_r$. The corresponding values for $\delta$ are calculated as for $|c|=d$.

In the latter two cases, two integers $r$ and $s$ enumerate different solution branches. From Eq.~\eref{eq:delta}, we see that increasing $s$ shifts the bifurcation lines towards larger values of $\delta$. By contrast, the influence of $r$ is less obvious. While different values of $r$ correspond to distinct bifurcation lines in the parameter space, solutions that differ only by $s$ often connect to each other such that they correspond to different segments of the same bifurcation line.

The results from the Sec.~\ref{BSS} and \ref{DelayOscillators} enable us to calculate the bifurcation points of instabilities that are created in BSS in which all C-nodes have an identical degree $d$. Even in large and complex networks instabilities that are generated entirely in certain symmetric subgraphs can thus be understood analytically.  

\section{Mesoscale induced instabilities}~\label{Numerics}
The results of the previous sections imply that a given BSS has a characteristic set of bifurcations that will appear in every network that contains the BSS. In the present section we support this conclusion with numerical evidence. 

\begin{figure}
\begin{center}
\includegraphics{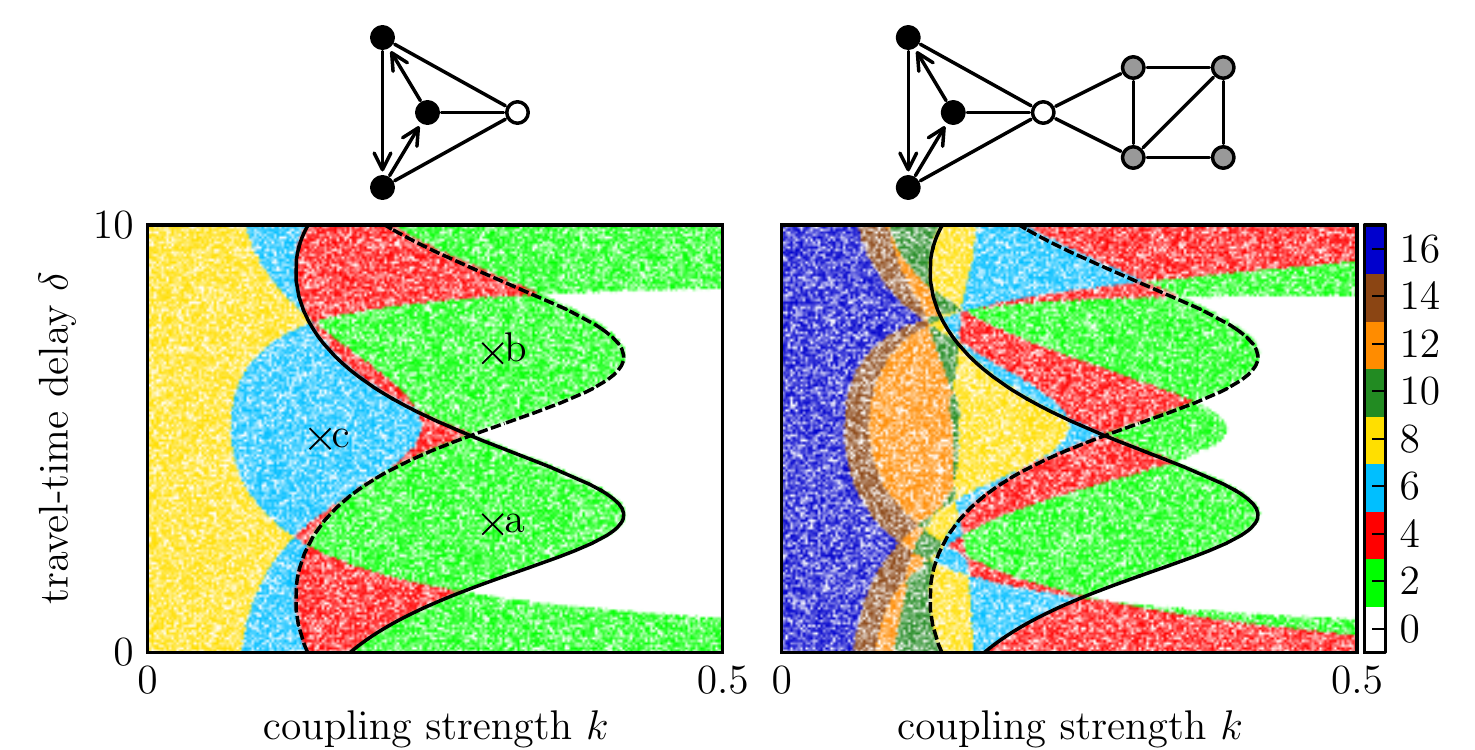}
\includegraphics{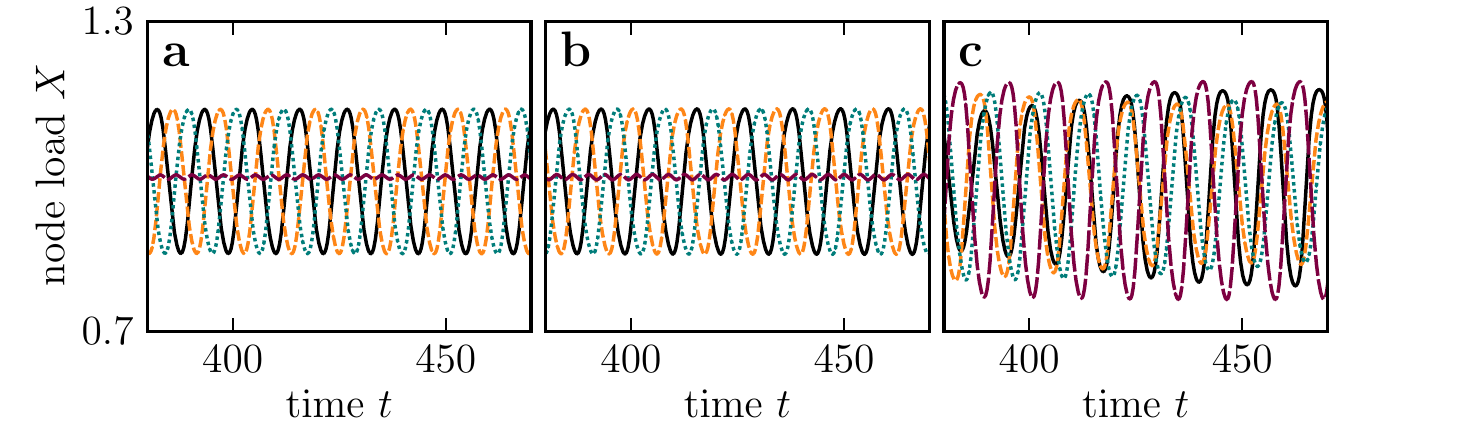}
\end{center}
\caption{Characteristic instabilities induced by the directed triangle subgraph. Top row: Diagrammatic representation of two network topologies, specifically the directed triangle with connecting nodes (left) and the a larger network containing the directed triangle subgraph. (Symbols as in Table 1). Center row: scatter plots showing the stability of specific pairs of the parameters ($\eps$, $\delta$). Colour coded is the number of dynamical eigenvalues with positive real parts, such that white corresponds to the region where the homogeneous steady state is stable. The solid and dashed lines correspond to analytical results for bifurcations that are localised in the triangle subgraph. The agreement between the scatter plot and the analytically obtained bifurcation lines show that the bifurcation lines of the two localised eigenvalues of the BSS are not affected by the residual graph. Bottom row: Time series from a delay-coupled Mackey-Glass systems with different parameters $\eps$ and $\delta$, marked by crosses in the scatter plots.
Colours correspond to different variables of the triangle subgraph, showing that the two localised instability correspond to right-handed and left-handed oscillations respectively.
Parameters are $\gp=-1.5, \lp=1, \eps=1$, $\tau=5$.} \label{fig:circle3}
\end{figure}

Because the computation of the potentially infinite set of eigenvalues is difficult, we use the approach proposed in \cite{Luzyanina1996,Johannes}, which uses Cauchy's argument principle to compute the number of eigenvalues with positive real parts. 
In the following we apply these approach to analyse networks with a given topology. For each of these networks we consider a large ensemble of ($\eps$,$\delta$) parameter pairs that are drawn independently from a uniform distribution. We show this ensemble as a scatter plot in the 
($\eps$,$\delta$)-plane, in which we colour-code the number of eigenvalues with positive real parts. As a result we obtain plots such as the ones shown in Fig.~\ref{fig:circle3}, where changes in colour indicate bifurcation lines.  

As a first example, we consider two networks containing the triangular BSS from Table~\ref{Examples}. 
For the directed triangle, $d=2$ as every inner node of the BSS has two incoming and two outgoing links. 
The bifurcation lines induced by the localised topological eigenvalues, $c=\exp(2\pi i/3)$ and $c=\exp(4\pi/3)$, can thus be calculated using Eq.~\eref{eq:AD_beta}. 
The scatter plot for the isolated BSS (Fig.~\ref{fig:circle3}, top left) shows a number of bifurcation lines. 
Some of these lines arise from delocalised instabilities. 
However, two of the bifurcation lines correspond to characteristic instabilities of the triangle. 
These lines agree exactly with the analytical results. 

Let us now consider the more complicated example shown on the right hand side of Fig.~\ref{fig:circle3} which also contains the directed triangle BSS.
For the larger network the corresponding scatter plot is significantly more complex. 
Note in particular that the delocalised instabilities that were present in the isolated triangle BSS do not carry over to the complex network. 
However, as we have shown analytically, the characteristic instabilities are independent of the embedding topology. 
Thus, the bifurcation lines corresponding to characteristic instabilities reappear exactly also in the larger network.

The effect of the characteristic instabilities on the system can be visualized by simulations (Fig.~\ref{fig:circle3}, bottom). In contrast to the results above, which are valid for the general class of models, simulations require restricting the functional form of the equations of motion. We use coupled Mackey-Glass systems  
\begin{equation}\label{Mackey_Glass}
\dot{X}_i=\frac{aX_i^\tau}{1+(X_i^\tau)^b}-c X_i+\epsilon \sum_j\left(A_{ij}X_j-A_{ji}X_i\right),
\end{equation}
with $a=2, b=5, c=1, \epsilon=10, \tau=5$ corresponding to the generalized parameters $\gp=-1.5, \lp=1, \eps=\epsilon \tau=1$. 
The simulation results show that the two characteristic instabilities correspond to clockwise and counter-clockwise oscillations on the triangle. One can therefore think of the directed triangle BSS as a network motor that will induce a characteristic forward or reverse oscillation if specific conditions are met, independently of the embedding topology. Let us emphasize that these oscillations do not generally remain confined to the triangle subgraph. After the onset of oscillations the propagation of the oscillatory solutions across the network is not described by the same Jacobian and the delicate balance which isolates the instability can be destroyed in the non-stationary state.  

\begin{figure}
\begin{center}
\includegraphics{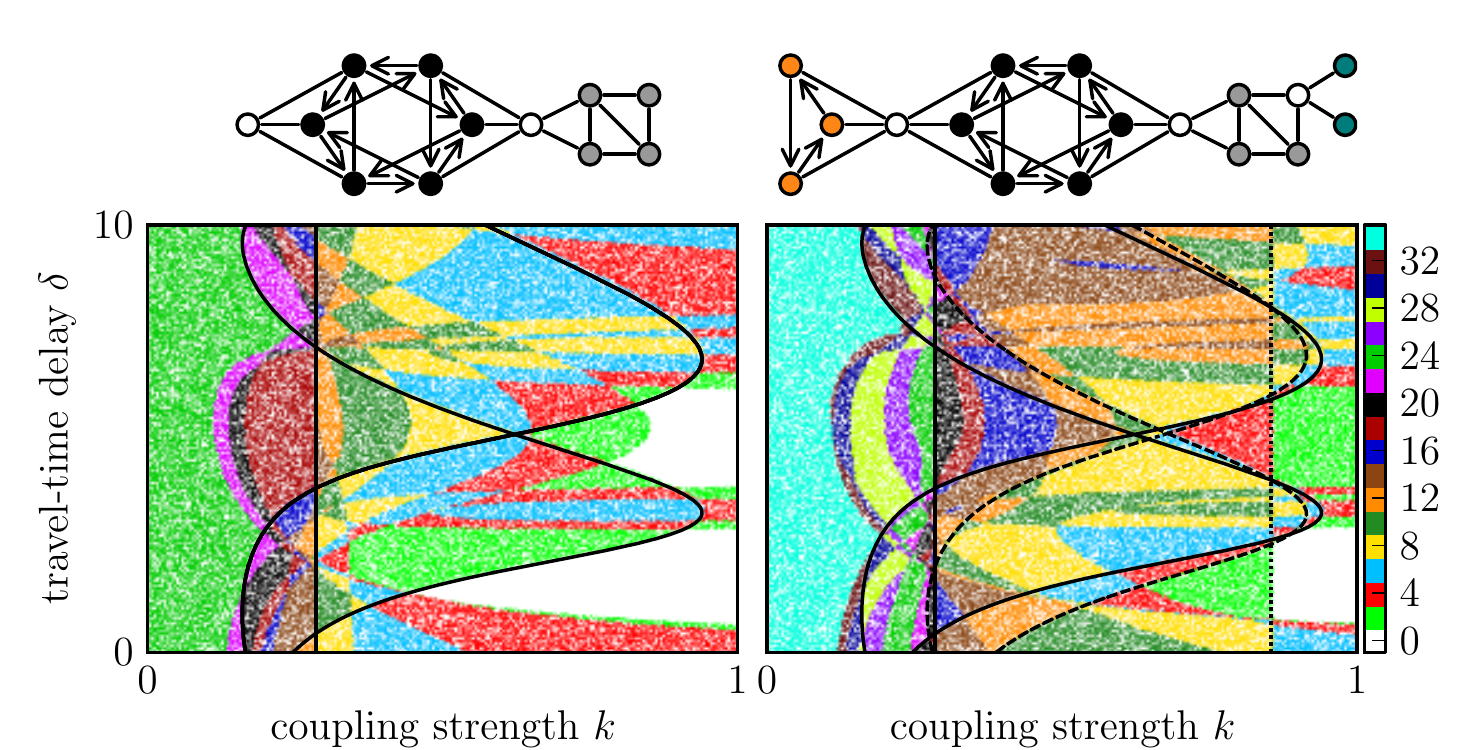}
\end{center}
\caption{Characteristic instabilities induced by different symmetric subgraphs. Plots are analogous to Fig.~\ref{fig:circle3}.
Both scatter plots contain the characteristic instabilities induced by the hexagonal BSS (solid black lines). Additionally, 
the network on the right-hand side contains also the characteristic instabilities of the directed triangle BSS (dashed, cf.~\ref{fig:circle3}) and the star-like BSS (dotted, vertical). For details on the corresponding localised eigenvalues see Table 1. This figure shows that characteristic instabilities induced by different BSS occurring in the same network superimpose without interfering.}\label{fig:complex}      
\end{figure}

Two more complicated examples are shown in Fig.~\ref{fig:complex}. The figure shows the scatter plots obtained for model realization on two different networks, which both contain the hexagonal BSS with two connecting nodes. The network on the right-hand side moreover contains the star-like BSS and the directed-triangle BSS. In both scatter plots, we find the characteristic bifurcation lines of the hexagonal BSS, corresponding to the topological eigenvalues $c=0$, $c=2\exp(2\pi i/3)$ and $c=2\exp(4\pi i /3)$. The two-fold degeneracy of the eigenvalue $c=0$ is reflected in the number of positive eigenvalues changing by 4 instead of 2 when crossing the corresponding bifurcation line (solid vertical line). In the scatter plot on the right, we further find the bifurcation lines of the triangular BSS (cf. Fig.~\ref{fig:circle3}) and the bifurcation line originating from the localised eigenvalue $c=0$ of the star-like BSS.
This confirms that bifurcations induced by different BSS can be superimposed without interfering.  

\section{Conclusions}~\label{Conclusions}
In the present paper we showed that symmetric subgraphs induce characteristic localised instabilities.
The bifurcations arising from these instabilities occur independently of the structure of the embedding network.  
These results have two implications. First they provide a wealth of examples of mesoscale structures that have precise and distinct implications for system-level dynamics. Second, because these implications are independent of the embedding structure, they can be analysed in small, structurally simple systems. Therefore, analytical solutions can be obtained even for relatively complicated dynamics such as the delay-coupled delay systems studied here.      

The beauty of the results lies perhaps in their simplicity. Essentially a symmetry in topology translates into a symmetry of topological eigenvalues in the complex plane, which then translates into a symmetry of dynamic eigenvalues, and onward to a corresponding dynamical mode. Here we have mainly considered the case of cyclic structures such as the directed triangle, which cause corresponding cyclic modes. It can be easily extrapolated that for instance bipartite structures caused characteristic instabilities that correspond to an imbalance between the partitions. For instance in a directed four-cycle, one would expect four cyclic instabilities and an additional instability where two nodes start to suppress the other two.    

Although the connection between mesoscale structure and dynamics may be intuitive, we note that it is only exactly true in the special case of exact symmetry that is not generic in the mathematical sense.
In a completely random system, one would not expect that any dynamical instability localises exactly on a finite set of nodes.  
It has been pointed out by \cite{McArthur} that symmetric structures are relatively common in real world networks, with up to 80 \% of the nodes belonging to symmetric structures in some examples. 
However, the symmetry of the Jacobian matrix on which we build here, requires additionally that the symmetric nodes also follow the same dynamical rules. 
In many biological systems this requirement is only be met exactly in special situations. 
Nevertheless, we expect the present results to remain valid to good approximation if the dynamical differences between topologically symmetric nodes are reasonably small (e.g. in a metapopulation where topologically symmetric nodes correspond to patches with similar but not identical environmental conditions). Furthermore, it is easily conceivable that almost exact symmetries exist in many technical systems. 
The exploration of both technological and biological implications of the present work, appear to be promising targets for future work, including the analysis of real-world data.  

Despite the caveat above, the connection between mesoscale structure and dynamics should remain valid in a wide variety of different systems. 
In the present paper we illustrated this connection by the example of delay-coupled delay oscillators. 
For simplicity we focused on topologies where the number of a node's outgoing links is identical to the number of incoming links.
Furthermore, we only considered symmetric subgraphs in which all inner nodes had the same number of links, such that the system was factorizing.
While these assumptions were essential for a clean illustration of the results, very similar results could have been obtained in other systems. First, we have focused here on delay-coupled delay oscillators, because they provide non-trivial examples, where already very simple structures can sustain complex dynamics. The same reasoning could have been presented for ordinary differential equations or discrete time maps, although perhaps slightly more complex symmetric subgraphs would have been needed to provide interesting examples. Second, the balance condition for links allowed us to avoid the computation of steady states which is impossible with the chosen degree of generality, but does not otherwise affect the subsequent reasoning. Third, the factorization enabled us to establish an explicit relationship between the topological and dynamical eigenvalues. We expect that the same will be possible for many symmetric structures and systems. However, even if this is not the case, a symmetry in the structure will generally induce a corresponding symmetry in the Jacobian if the symmetric nodes follow the same dynamical rules. Thus, even when the connection between structure and dynamics cannot be made explicit, a structural symmetry in a subgraph will generally induce characteristic instabilities that are independent of the embedding network. Because these instabilities originate in small sub-systems their numerical investigation should in general be possible with relative ease (see \cite{Helge} for an example).    
 
One difficulty that may be encountered in future work is that the Jacobian can be sensitive to the location of the steady state. 
While an embedding network cannot affect the characteristic bifurcation lines in the bifurcation diagrams shown here, it can, by shifting the steady state, alter the parameter values at which the system is located within the bifurcation diagram. 
Our results would thus have been less clear if did not use the generalised modelling formalism that conceptually separates the effects of shifting steady state concentrations from the effects of changes in the nonlinearities \cite{Gross}.  
In practice this means that subgraphs, such as the directed-triangle, can be constructed to sense some information from the surrounding network if this information is communicated via a change in the steady state values of the variables associated with the symmetric nodes.   
Conversely, one can also construct networks that are completely insensitive to the surrounding network.
This can be done for instance by choosing the functional forms of the processes in the model to follow power laws. For such functions the generalized parameters that appear in the Jacobian are independent of the steady state under consideration (cf.~Eq.~(\ref{eqGenPar})).   
      
In conclusion, the work presented here points to a general connection between the structure of certain mesoscale subgraphs and system-level dynamics. It is conceivable that, in future works, this relationship may prove instrumental for understanding or engineering certain dynamics in complex nonlinear systems. For instance our results seem to suggest that networks can be constructed such that some function that is sustained on a certain part of the network can be isolated from functions that are localised on other parts of the net. Such localization is indeed observed in many biological networks, but significantly more work will be necessary to determine whether the symmetries described here contribute to this localization. Furthermore, attempts could be made to engineer simple information processing systems based on communicating symmetric subgraphs that are embedded in a larger network. While much work remains to be done to explore these possibilities we believe that, if successful, these efforts might lead to design principles for information processing in applications such as synthetic biology or chemical computing.

\end{document}